\documentclass[
               biblatex,     
               ]{jacow}

\usepackage[english]{babel}
\begin{document}

\title{Dispersion Suppression for Wedge-Based Final Cooling at a 10 T\MakeLowercase{e}V Muon Collider\thanks{Endorsed by the International Muon Collider Collaboration.}}

\author{
I. Karaaslan\thanks{inci@uchicago.edu}, K. DiPetrillo, 
University of Chicago, Chicago, IL, USA\\
D. Neuffer, D. Stratakis, K. Yonehara, 
Fermi National Accelerator Laboratory, Batavia, IL, USA
}

\maketitle

\begin{abstract}
	Achieving a luminosity of $\gtrsim\qty{e34}{\cm\tothe{-2} \s\tothe{-1}}$ in a $\qty{10}{TeV}$ Muon Collider, given the short lifetime of a muon, requires reducing the 6D emittance of the muon beam through a process known as ionization cooling. In the final stage of this cooling process, the transverse emittance must be reduced to $\sim\qty{22}{\micro\metre}$, typically by allowing longitudinal emittance growth up to downstream acceptance limits. While the current International Muon Collider Collaboration designs involve $\qty{40}{T}$ solenoids to reach the transverse emittance target, such high-field solenoids come with several challenges, including mechanical stress management, quench protection, and potential limitations in relying on High Temperature Superconductor technology. Designed as an alternative to using such solenoids while simultaneously reaching target transverse emittance, the previously proposed wedge-based, reverse emittance-exchange cooling scheme requires excellent dispersion suppression. In this study, we design and simulate a dispersion suppressor channel for the wedge-based final cooling design that reduces dispersion in the target direction to a target value of $D_x \sim \qty{0.001}{\m}$. 
\end{abstract}

\section{INTRODUCTION}

\begin{figure}[!htb]
   \centering
   \includegraphics*[width=0.8\columnwidth]{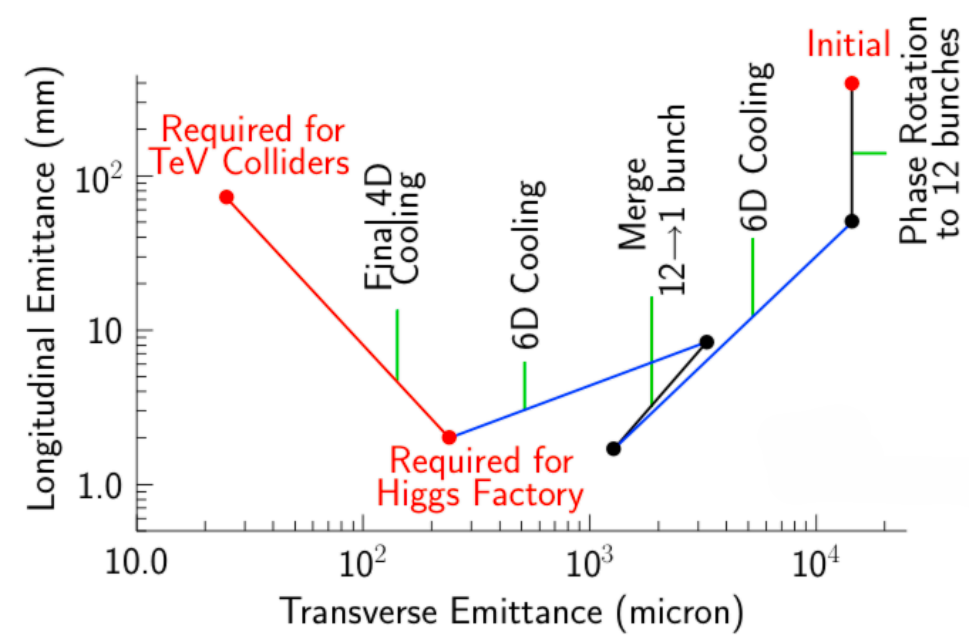}
   \caption{Fernow-Neuffer plot of the muon cooling process~\cite{yonehara:ipac10-mopd076, yonehara:cool25-2025}. The Final 4D Cooling stage investigated in this study is shown in a red line.}
   \label{fig:1}
\end{figure}

In the realm of planned next-generation facilities following the completion of the High Luminosity Large Hadron Collider (HL-LHC) at CERN, a circular muon collider stands out from others due to its potential to reach a multi-TeV center of mass energy with greater energy and power efficiency. For a feasible realization of the International Muon Collider Collaboration (IMCC) $\qty{10}{TeV}$ muon collider design, obtaining luminosities of $\gtrsim\qty{e34}{\cm\tothe{-2} \s\tothe{-1}}$ at the interaction point through emittance reduction at the cooling stage of the collider is crucial~\cite{schulte:ipac22-tuizsp2,schulte:ipac21-thpab017}. Fig.~\ref{fig:1} shows the evolution of the muon beam's transverse and longitudinal emittance throughout all cooling stages. The IMCC baseline Final Cooling design features liquid hydrogen ($LH_2$) absorbers and RF cavities within high-field matching solenoids. In more recent designs, the field strength of these solenoids gets up to $\geq \qty{40}{T}$~\cite{fabbri:ipac23-wepm062} at the absorbers. 

Creating and maintaining this ultra-high field regime is one of the major challenges of the design, as such magnets introduce large electromagnetic stresses, the need for quench protection and thermal management (a summary of further progress and challenges can be found in \cite{bermudez_snowmass_nodate}). In this paper, we outline a previously proposed alternative final cooling concept that doesn't rely on high-field solenoids, discuss the requirements for the realization of the full design with a focus on dispersion suppression, and finally demonstrate dispersion suppression with our design.

\section{WEDGE-BASED FINAL COOLING \\ DESIGN OVERVIEW}
\begin{figure}[htb]
   \centering
   \includegraphics*[width=\columnwidth, height=3 cm]{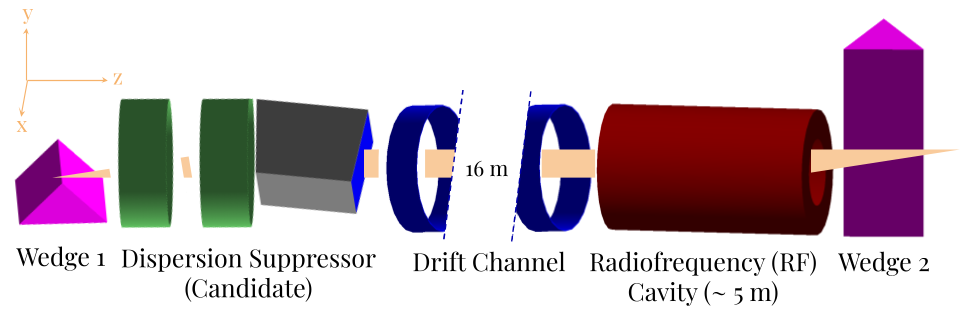}
   \caption{Wedge-based final cooling lattice modeled in G4Beamline with the candidate dispersion suppressor design (two quadrupoles and a sector dipole). The orange band indicates the beam trajectory along z. The first wedge is oriented in the x direction, whilst the second wedge at the end is oriented towards the y direction, facilitating ionization cooling in x and y directions respectively.}
   \label{fig:2}
\end{figure}

The optimized wedge-based final cooling design is based on the concept outlined in~\cite{neuffer:ipac15-tubd2,neuffer:ipac24-tups20}. A diagram of the updated design can be found in Fig.~\ref{fig:2}. We consider a starting distribution of $\qty{145}{\mu m}$ transverse emittance with a momentum spread of $\sigma_p = \qty{1.0}{MeV/c}$, corresponding to stage B10 of rectilinear 6D ionization cooling design (originally ~\cite{PhysRevSTAB.18.031003} involved 8 stages, but more recent improvements on the original design involve stage B10~\cite{Summers2020}). The muon beam is initialized with a mean momentum of $p_\mu = \qty{100}{MeV/c}$ where an emittance exchange can be realized through a single wedge-shaped solid "dense carbon" diamond material~\cite{neuffer:ipac24-tups20} (referred to as just "wedge" hereafter). Dispersion is introduced in the wedge through the wedge's position-dependent non-uniform thickness, exploiting existing momentum-position correlation to preferentially reduce the longitudinal emittance while increasing transverse emittance, achieving reverse-emittance exchange.

The magnets in the dispersion suppression section succeeding the first wedge are positioned to primarily mitigate the dispersion in the x direction. This dispersion correction section design is composed of elements whose fields are linear in transverse coordinates, i.e., dipoles and quadrupoles, which also focus off-momentum particles to the centerline. Prior to this study, the effects of this dispersion section for the final cooling channel have been shown only through manual subtraction of the dispersion from the post-wedge particle distribution.

After the dispersion correction channel, the beam goes through a drift channel to increase the bunch length $\sigma_t$ so that the subsequent radiofrequency (RF) cavity can do an energy-phase rotation to reduce $\sigma_p$, the momentum spread of the particle distribution. Any residual dispersion uncorrected by the dispersion suppressor, which corresponds to a nonzero correlation between transverse position and relative momentum, will cause coupling between transverse and longitudinal planes in the beam line, reducing the effectiveness of the energy-time phase rotation. The first part with the wedge and the dispersion suppression section could be repeated for the y-axis, thus ensuring a complete emittance reduction of all transverse directions. 

\section{DISPERSION SUPPRESSION DESIGN}
\begin{table}[htb]
   \centering
   \caption{Relevant Parameter Values for the Initial Beam After the First Wedge}
   \begin{tabular}{lcc}
       \toprule
       \textbf{Name} & \textbf{x Axis} & \textbf{y Axis} \\ 
       \midrule
          $\epsilon$ [mm]    & \num{0.042}       & \num{0.146}        \\
          $\beta$ [m]  & \num{0.035}       & \num{0.025}        \\
          $\alpha$ & \num{-2.437}       & \num{-0.681}        \\
          D [m]  & \num{0.01989}       & \num{-0.0002}        \\
          D'  & \num{-0.0402}       & \num{-0.0128}        \\
       \bottomrule
   \end{tabular}
   \label{tab:1}
\end{table}

In order to discuss the optimized design candidate for dispersion suppression, we first motivate the target values for such a design. Table~\ref{tab:1} shows the relevant Twiss parameters for the muon beam after traversing through the wedge with wedge length of $\qty{7.51}{\mm}$ and wedge angle of $\qty{47.3}{^\circ}$ and a mean momentum of $p_\mu = \qty{87.33}{MeV\per c}$ and a spread of $\sigma_p = \qty{7.15}{MeV/c}$. In dispersive systems, we can write the RMS beam size as:
\begin{equation}\label{eq:1}
    \sigma_x^2 = \epsilon_x \beta_x + D_x^2\sigma_\delta^2
\end{equation} where $\sigma_\delta = \Delta p/p_\mu$~\cite{Edwards:1992unz}. Simply inputting the values from the table, we can observe that the second dispersive term in Eq. ~\ref{eq:1} is higher than the first geometric term, indicating that the post-wedge beam size that will be propagated downstream the cooling channel is affected by dispersion. For the beam size to be less affected, we render the dispersive contribution subdominant by targeting $D_x^2\sigma_\delta^2 \lesssim 10\%$ $\epsilon_x \beta_x$. This means that the dispersion value targeted after the dispersion suppressor must be at least one order of magnitude lower than its post-wedge value.

\subsection{Optimization Method}
\begin{table}[htb]
   \centering
   \caption{Relevant Parameter Values for the Beam After QQB}
   \begin{tabular}{lcc}
       \toprule
       \textbf{Name} & \textbf{x Axis} & \textbf{y Axis} \\ 
       \midrule
          $\epsilon$ [mm]    & \num{0.337}       & \num{0.148}        \\
          $\beta$ [m]        & \num{1.25}       & \num{5.31}         \\
          $\alpha$           & \num{-2.01}      & \num{-24.69}       \\
          $D$ [m]            & \num{0.0009}    & \num{-0.00076}       \\
          $D'$               & \num{0.0103}       & \num{-0.0028}        \\
          \midrule
          Transmission After QQ [\%] & \multicolumn{2}{c}{\num{93.5}} \\
          Transmission After B  [\%]  & \multicolumn{2}{c}{\num{84.6}} \\
       \bottomrule
   \end{tabular}
   \label{tab:2}
\end{table}
\begin{table}[htb]
   \centering
   \caption{Optimized QQB Parameters}
   \begin{tabular}{lcc}
       \toprule
       \textbf{Parameter} & \textbf{Value} \\ 
       \midrule
          QF\_gradient [T/m]   & \num{4.47} \\
          QD\_gradient [T/m]   & \num{6.68} \\
          Q\_radius [mm]  & \num{71.41} \\
          Drift\_length [mm]  & \num{10.00} \\
          Q\_z [mm] & \num{107.9}\\
          B\_field [T]  & \num{1.74}\\
          B\_angle [deg]  & \num{1.63}  \\
          B\_bending\_radius [mm]  & \num{643.1}  \\
          B\_z [mm]  & \num{5.99}  \\
       \bottomrule
   \end{tabular}
   \label{tab:3}
\end{table}
\begin{figure}[htb]
   \centering
   \includegraphics*[width=\columnwidth, height=4.5cm]{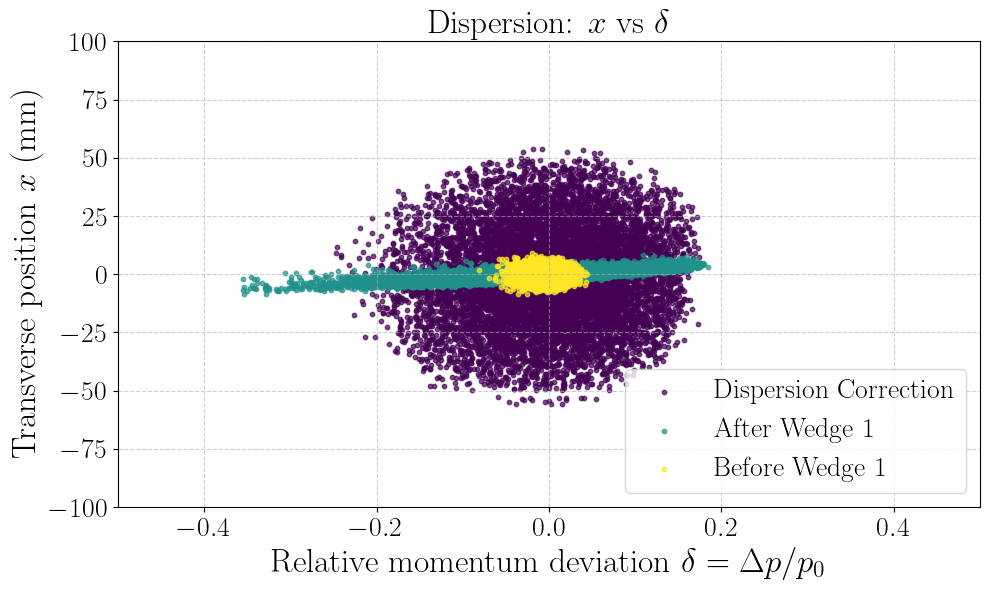}
   \caption{The dispersion vs. $\delta$ plot for various stages: before the wedge, after the wedge, and after the dispersion suppressor (with a 2$\sigma$ cut applied).}
   \label{fig:3}
\end{figure}
To guide the design of the dispersion suppressor, we first consider established approaches such as the Double Bend Achromat (DBA) lattice as in ~\cite{chasman:pac75} and the FODO-based dispersion suppression scheme as in ~\cite{Edwards:1992unz}. The DBA lattice consists of a focusing quadrupole magnet located halfway between a pair of identical dipole magnets, and is typically constructed to enforce $D=D'=0$ at its entrance and exit. These conditions are not satisfied in our case, where the beam exiting the wedge carries nonzero $D, D'$. Similarly, FODO-based dispersion suppressors assume the beam comes from a repeating sequence of focusing and bending elements that make up the arc of a storage ring or recirculating accelerator, where dispersion is generated and subsequently canceled by modifying the bending distribution. This makes the design incompatible with our single-pass cooling channel with an arbitrary $D, D'$ from the wedge, as the periodicity required to analytically cancel dispersion is not present. 

We instead adopt a more flexible approach, comprising a quadrupole doublet and a sector dipole (QQB), where the quadrupoles provide transverse focusing and matching and the dipole corrects the residual beam dispersion. To design and refine this QQB lattice, we created two G4Beamline templates: one for the focusing (\verb|QF|) and defocusing (\verb|QD|) quadrupoles separated by a drift channel, and another for the sector dipole. A companion Python script performs optimization over the parameters \verb|QF_gradient, QD_gradient, Drift_length| for the quadrupole doublet (representing the focusing and defocusing quadrupole gradients, the length of the drift channel between the quadrupoles, respectively) and \verb|B_strength, B_angle, B_bending_radius| for the dipole (defining field strength, sector angle, radius, respectively). \verb|Q_z, B_z| are also included, as they describe the distance to the initialized beam from the first quadrupole and the sector dipole, respectively. Quadrupole radii and dipole bending radius boundaries are determined based on the incoming beam size to maximize transmission. 

\begin{figure}[htb]
   \centering
   \includegraphics*[width=0.95\columnwidth]{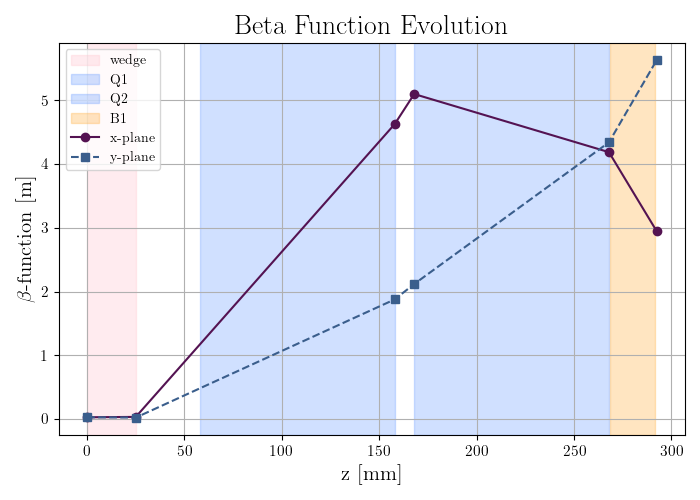}
   \includegraphics*[width=0.95\columnwidth]{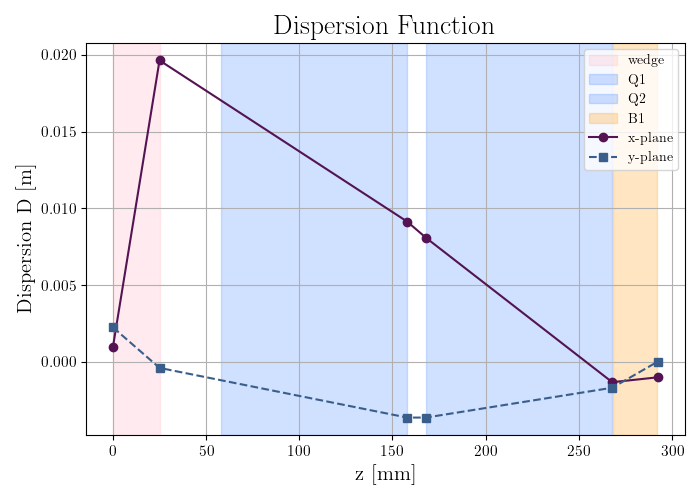}
   \includegraphics*[width=0.95\columnwidth]{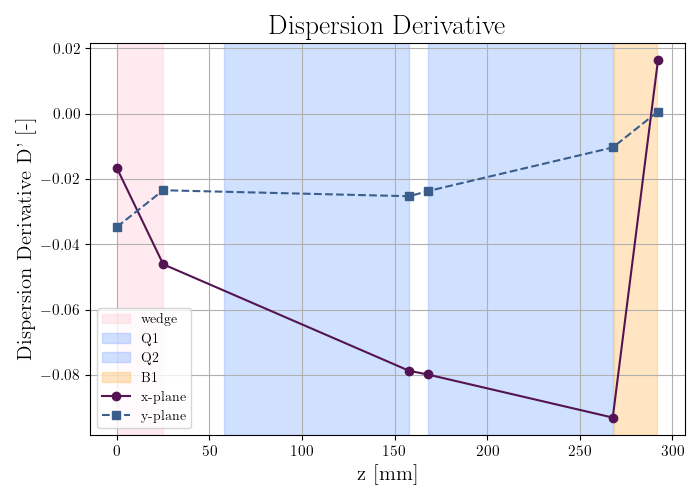}
   \caption{The evolution of Twiss parameters in the QQB dispersion suppressor with respect to the longitudinal beam axis.}
   \label{fig:4}
\end{figure}

The optimization proceeds in two stages: we first optimize the quadrupole parameters using a global Differential Evolution search~\cite{Storn:1997uea}, \cite{2020SciPy-NMeth}, followed by local refinement with the Nelder-Mead algorithm~\cite{journals/cj/NelderM65}. The resulting particle distribution is then fed into the dipole, where dipole parameters are optimized similarly. During this step, the beta penalties are weighted more heavily for quadrupoles, while the dispersion penalty receives a higher weight for the dipole. Table~\ref{tab:2} shows the relevant Twiss parameters for the beam after going through the optimized QQB system. The transverse phase space evolution from the wedge to the end of the dispersion suppressor can be seen in Fig.~\ref{fig:3}, and the evolution of these parameters throughout the system is illustrated in Fig.~\ref{fig:4}. Table~\ref{tab:3} shows the values of the optimized parameters for the QQB design.

\subsection{Discussion and Future Work}
Comparing the values in Table~\ref{tab:2} with the target estimates discussed previously, we find that the target suppression of $D_x, D_y$ has been achieved. However, this comes with a clear trade-off: while the initial particle distribution exhibits low beta functions, the stronger focusing required to reduce dispersion leads to an increase in $\beta_x, \beta_y$ as well as an increase in transverse emittance in the x direction. We also note a trend of increasing beta function values with momentum spread $\delta$, which suggests the presence of chromatic effects induced by the quadrupoles. The growing transverse emittance in the x direction for the dispersion correction design could point toward the growth of stochastic effects introduced by the wedge, which could be causing deviations from ideal linear optics. Although these effects are not explicitly quantified here, they may become relevant for downstream beam dynamics and warrant further investigation. Similar designs, such as BQQ or BQQB, prove to have a similar potential, so more work on exploring all potential designs would be needed. Future designs should look to combine this dispersion suppressor with the rest of the downstream elements of the final cooling channel to quantify the effect of dispersion on the final transverse emittance value.

\section{CONCLUSION}
We have designed and developed parameters for a dispersion suppressor in a wedge-based final cooling system for a Muon Collider. Directions for future development are discussed.

\newpage \newpage
\printbibliography

@inproceedings{schulte:ipac21-thpab017,
    author = {D. Schulte},
    title = {{The International Muon Collider Collaboration}},
    booktitle = {Proc. IPAC'21},
    pages = {3792--3795},
    paper = {THPAB017},
    year = {2021},
    venue = {Campinas, Brazil, May 2021},
    publisher = {JACoW Publishing, Geneva, Switzerland},
    doi = {10.18429/JACoW-IPAC2021-THPAB017},
    url = {https://jacow.org/ipac2021/papers/THPAB017.pdf},
    language = {english}
}

@inproceedings{schulte:ipac22-tuizsp2,
    author = {D. Schulte},
    title = {{The Muon Collider}},
    booktitle = {Proc. IPAC'22},
    pages = {821--826},
    paper = {TUIZSP2},
    venue = {Bangkok, Thailand},
    series = {International Particle Accelerator Conference},
    number = {13},
    publisher = {JACoW Publishing, Geneva, Switzerland},
    month = {7},
    year = {2022},
    issn = {2673-5490},
    isbn = {978-3-95-450227-1},
    doi = {10.18429/JACoW-IPAC2022-TUIZSP2},
    url = {https://jacow.org/IPAC2022/papers/TUIZSP2.pdf},
    language = {english}
}

@inproceedings{fabbri:ipac23-wepm062,
    author = {S. Fabbri and others},
    title = {{Magnets for a muon collider}},
    booktitle = {Proc. IPAC'23},
    pages = {3712--3715},
    paper = {WEPM062},
    venue = {Venice, Italy},
    series = {International Particle Accelerator Conference},
    number = {14},
    publisher = {JACoW Publishing, Geneva, Switzerland},
    month = {9},
    year = {2023},
    issn = {2673-5490},
    isbn = {978-3-95-450231-8},
    doi = {10.18429/JACoW-IPAC2023-WEPM062},
    language = {english}
}

@misc{bermudez_snowmass_nodate,
  author = {Bermudez, S.I. and Sabbi, G. and Zlobin, A.},
  title = {Snowmass Accelerator Conveners (AF1-AF7, ITF, ee/mmFora) Meeting \#17},
  url = {https://indico.fnal.gov/event/54953/sessions/20614/attachments/156153/205927/2022-07-18_Snowmass_Summary_AF7Magnets_draft.pdf},
  urldate = {2026-03-11},
  journal = {INDICO-FNAL (Indico)},
  %note = {this meeting},
  key = {snowmass2022}
}

@inproceedings{yonehara:ipac10-mopd076,
    author = {K. Yonehara and Y. S. Derbenev and R. P. Johnson and M. L. Neubauer},
    title = {{A Helical Cooling Channel System for Muon Colliders}},
    booktitle = {Proc. IPAC'10},
    pages = {870--872},
    paper = {MOPD076},
    year = {2010},
    venue = {Kyoto, Japan, May 2010},
    publisher = {JACoW Publishing, Geneva, Switzerland},
    url = {http://accelconf.web.cern.ch/IPAC10/papers/MOPD076.pdf},
    language = {english}
}

@misc{yonehara:cool25-2025,
  author = {Yonehara, Katsuya},
  title = {Collective Effects in Muon Ionization Cooling},
  year = {2025},
  doi = {10.2172/3018551},
  %url = {https://www.osti.gov/servlets/purl/3018551},
  note = {presented at the COOL'25 Workshop on Beam Cooling and Related Topics},
  %urldate = {2026-04-01}
}

@inproceedings{neuffer:ipac15-tubd2,
    author = {D. V. Neuffer and T. L. Hart and D. J. Summers and H. K. Sayed},
    title = {{Final Cooling For a High-luminosity High-energy Lepton Collider}},
    booktitle = {Proc. IPAC'15},
    pages = {1384--1386},
    paper = {TUBD2},
    year= {2015},
    venue = {Richmond, VA, USA, May 2015},
    publisher = {JACoW Publishing, Geneva, Switzerland},
    doi = {10.18429/JACoW-IPAC2015-TUBD2},
    url = {https://jacow.org/IPAC2015/papers/TUBD2.pdf},
    language = {english}
}

@inproceedings{neuffer:ipac24-tups20,
    author = {D. Neuffer and D. Stratakis},
    title = {{Final cooling with thick wedges for a muon collider}},
    booktitle = {Proc. IPAC'24},
    pages = {1684--1686},
    paper = {TUPS20},
    venue = {Nashville, TN, USA},
    series = {International Particle Accelerator Conference},
    number = {15},
    publisher = {JACoW Publishing, Geneva, Switzerland},
    month = {7},
    year = {2024},
    issn = {2673-5490},
    isbn = {978-3-95-450247-9},
    doi = {10.18429/JACoW-IPAC2024-TUPS20},
    language = {english}
}

@article{PhysRevSTAB.18.031003,
  title = {Rectilinear six-dimensional ionization cooling channel for a muon collider: A theoretical and numerical study},
  author = {Stratakis, Diktys and Palmer, Robert B.},
  journal = {Phys. Rev. ST Accel. Beams},
  volume = {18},
  issue = {3},
  pages = {031003},
  numpages = {11},
  year = {2015},
  month = {Mar},
  publisher = {American Physical Society},
  doi = {10.1103/PhysRevSTAB.18.031003},
  url = {https://link.aps.org/doi/10.1103/PhysRevSTAB.18.031003}
}

@book{Edwards:1992unz,
    author = "Edwards, D. A. and Syphers, M. J.",
    title = "{An Introduction to the Physics of High-Energy Accelerators}",
    isbn = "978-0-471-55163-8",
    publisher = "Wiley",
    address = "New York",
    series = "Wiley Series in Beam Physics and Accelerator Technology",
    year = "1992"
}

@inproceedings{chasman:pac75,
    author = {R. Chasman and G. K. Green and E. M. Rowe},
    title = {{Preliminary Design of a Dedicated Synchrotron Radiation Facility}},
    booktitle = {Proc. PAC'75},
    pages = {1765--1768},
    paper = {},
    year={1975},
    venue = {Washington D.C., USA, Mar. 1975},
    publisher = {JACoW Publishing, Geneva, Switzerland},
    language = {english}
}

@ARTICLE{2020SciPy-NMeth,
  author  = {Virtanen, Pauli and Gommers, Ralf and Oliphant, Travis E. and
            Haberland, Matt and Reddy, Tyler and Cournapeau, David and
            Burovski, Evgeni and Peterson, Pearu and Weckesser, Warren and
            Bright, Jonathan and {van der Walt}, St{\'e}fan J. and
            Brett, Matthew and Wilson, Joshua and Millman, K. Jarrod and
            Mayorov, Nikolay and Nelson, Andrew R. J. and Jones, Eric and
            Kern, Robert and Larson, Eric and Carey, C J and
            Polat, {\.I}lhan and Feng, Yu and Moore, Eric W. and
            {VanderPlas}, Jake and Laxalde, Denis and Perktold, Josef and
            Cimrman, Robert and Henriksen, Ian and Quintero, E. A. and
            Harris, Charles R. and Archibald, Anne M. and
            Ribeiro, Ant{\^o}nio H. and Pedregosa, Fabian and
            {van Mulbregt}, Paul and {SciPy 1.0 Contributors}},
  title   = {{{SciPy} 1.0: Fundamental Algorithms for Scientific
            Computing in Python}},
  journal = {Nature Methods},
  year    = {2020},
  volume  = {17},
  pages   = {261--272},
  adsurl  = {https://rdcu.be/b08Wh},
  doi     = {10.1038/s41592-019-0686-2},
}

@article{Storn:1997uea,
    author = "Storn, Rainer and Price, Kenneth",
    title = "{Differential Evolution {\textendash} A Simple and Efficient Heuristic for global Optimization over Continuous Spaces}",
    doi = "10.1023/A:1008202821328",
    journal = "J. Global Optim.",
    volume = "11",
    number = "4",
    pages = "341--359",
    year = "1997"
}

@article{journals/cj/NelderM65,
  added-at = {2019-07-23T00:00:00.000+0200},
  author = {Nelder, John A. and Mead, R.},
  biburl = {https://www.bibsonomy.org/bibtex/20ef42fec844fc18115f072b3191f41f1/dblp},
  ee = {https://www.wikidata.org/entity/Q55954356},
  interhash = {a5e0b861cf4f8ed6e67e8ea7cdc4b9ff},
  intrahash = {0ef42fec844fc18115f072b3191f41f1},
  journal = {Comput. J.},
  keywords = {dblp},
  number = 4,
  pages = {308-313},
  timestamp = {2019-07-24T11:36:54.000+0200},
  title = {A Simplex Method for Function Minimization},
  doi = {https://doi.org/10.1093/comjnl/8.1.27},
  %url = {http://dblp.uni-trier.de/db/journals/cj/cj7.html#NelderM65},
  volume = 7,
  year = 1965
}

@misc{Summers2020,
  author       = {D. Summers},
  title        = {More Muon Cooling, Higher Luminosity},
  year         = {2020},
  month        = nov,
  note         = {Presentation at the Muon Capture and Cooling Working Group, CERN, 12 Nov 2020},
  url          = {https://indico.cern.ch/event/961803/contributions/4064621/attachments/2141709/3608895/DS_mu_12Nov.pdf}
}

\end{document}